\documentclass[reqno]{amsart}
\usepackage{graphicx, amssymb, amsmath, latexsym}
\usepackage[utf8]{inputenc}
\usepackage{cite}
\usepackage{float}
\usepackage[margin=2cm]{geometry}
{ 
\begin{document}

\title{Continuum modeling perspectives of non-Fourier heat conduction in biological systems}

\author{Á. Sudár$^{1,2}$, G. Futaki$^{1}$, R. Kovács$^{1,2,3}$}

\address{
$^1$Department of Energy Engineering, Faculty of Mechanical Engineering, BME, Budapest, Hungary\\
$^2$Department of Theoretical Physics, Wigner Research Centre for Physics,
Institute for Particle and Nuclear Physics, Budapest, Hungary \\
$^3$Montavid Thermodynamic Research Group
}
\vspace{10pt}

\maketitle

\begin{abstract}
The thermal modeling of biological systems has increasing importance in developing more advanced, more precise techniques such as ultrasound surgery. One of the primary barriers is the complexity of biological materials: the geometrical, structural, and material properties vary in a wide range, and they depend on many factors. Despite these difficulties, there is a tremendous effort to develop a reliable and implementable thermal model.
In the present paper, we focus on the continuum modeling of heterogeneous materials with biological origin. There are numerous examples in the literature for non-Fourier thermal models. However, as we realized, they are associated with a few common misconceptions.
Therefore, we first aim to clarify the basic concepts of non-Fourier thermal models. These concepts are demonstrated by revisiting two experiments from the literature in which the Cattaneo-Vernotte and the dual phase lag models are utilized. Our investigation revealed that using these non-Fourier models is based on misinterpretations of the measured data, and the seeming deviation from Fourier's law originates in the source terms and boundary conditions.
\end{abstract}

%
\vspace{2pc}

\section{Introduction}
Thermodynamics is one of the most crucial fundamental blocks of interdisciplinary researches. It is proved to be useful in numerous coupled problems such as thermo-electricity, thermo-diffusion, and thermo-acoustics \cite{GrooMaz63non, Onsager31I, Onsager31II}, thanks to the universal definition of the thermodynamic laws. Regarding the topic in this paper, we consider only the I.~and the II.~laws of thermodynamics for heat conduction in a continuum medium. Moreover, we also assume the material to be rigid, although it is not characteristic of biological systems. We want to keep the discussion as simple as possible to clearly present the essential aspects of non-Fourier heat conduction for such complex systems. Otherwise, mechanics must be included together with its consequences: including source and coupling terms in the balance and constitutive equations, respectively.

These equations serve the basis for continuum models, i.e., the I.~law of thermodynamics describes the balance of internal energy:
\begin{equation}
\rho \partial_t e + \nabla \cdot \mathbf q = 0, \label{eq1}
\end{equation}
with $\mathbf q$ being the heat flux, $\rho$ is the mass density, $\nabla$ is the usual nabla operator, $e=c T$ is the internal energy density with $c$ standing for the specific heat, and $T$ is the temperature. Also, the zero on the right-hand side in (\ref{eq1}) expresses the lack of source terms. This equation requires a `closure,' an expression between the temperature $T$ and the heat flux $\mathbf q$. Such expression is called constitutive equation, describing a material's behavior, and strictly restricted by the II.~law of thermodynamics \cite{Mato04b, BerVan17b}. In the simplest model, one can use Fourier's law:
\begin{equation}
\mathbf q =- \lambda \nabla T, \label{eq2}
\end{equation}
in which $\lambda$ is the thermal conductivity. However, as observed in many cases \cite{JacWalMcN70, Naretal75, McN74t}, this relation is restricted on a certain time, and spatial scale \cite{JozsKov20b}, and can lose its validity. Therefore, one must find a suitable, thermodynamically compatible generalization.

One often referred example in papers on bioheat transfer is the dual phase lag (DPL) model,
\begin{equation}
\mathbf q(x, t+\tau_q) = -\lambda \nabla T(x,t+\tau_T), \label{eq3}
\end{equation}
introduced by Tzou \cite{Tzou95}. Here, $\tau_q$ and $\tau_T$ are the corresponding relaxation times (time lags), vaguely described in many papers. According to the original argument of Tzou, $\tau_q$ can be either smaller or higher than $\tau_T$, depending on the phenomenon. Sadly, numerous works concluded that these relaxation times are not independent of each other and cannot be arbitrary. Furthermore, the DPL model can violate the basic physical principles \cite{FabLaz14a, Quin07, KovVan18dpl, Ruk14, Ruk17}, depending on $\tau_q$ and $\tau_T$. This problem originates in the Taylor series expansion of equation (\ref{eq3}), considered up to arbitrary orders, which is not a thermodynamically compatible way to generalize and derive constitutive relations.

Instead of the DPL model, we propose a similar one with stronger physical background, called Guyer-Krumhansl (GK) equation. In what follows, we first shortly gather the thermodynamic background of the GK model. After, we briefly summarize the known experiences on non-Fourier heat conduction experiments, restricting ourselves on heterogeneous macro-scale materials and room temperature conditions. Then, we revisit two experiments in which both Fourier and non-Fourier models are utilized, showing common misconceptions about the possible non-Fourier effects. With revisiting these experiments, we aim to highlight the less-known properties of non-Fourier models and present our perspectives of modeling heterogeneous materials.

\section{Thermodynamics of constitutive relations}
As we previously mentioned, the II.~law of thermodynamics restricts the constitutive equations. In fact, it offers a constructive methodology to find a `closure' and derive the possible set of constitutive equations. The framework of Classical Irreversible Thermodynamics \cite{GrooMaz63non, Lebon89} formulate the entropy density $s$ as a potential function of the state variables. For heat conduction, it reads $s=s(e)$, and the Gibbs relation $\textrm{d}e=T\textrm{d}s$ expresses the functional relationship between $e$ and $s$. The II.~law can be formulated similarly to the I.~law:
\begin{equation}
\rho \partial_t s + \nabla \cdot \mathbf J_s = \sigma_s \geq 0, \label{eq4}
\end{equation}
where $ \mathbf J_s$ stands for the entropy flux, and $\sigma_s$ is the entropy production. Eventually, that inequality leads to the set of constitutive equations. `Set' is intentional: there are infinitely many solutions of such inequality. For instance, in case of $s=s(e)$ with $\mathbf J_s=\mathbf q/T$, it yields
\begin{equation}
\sigma_s = \mathbf q \nabla \frac{1}{T} \geq 0, \label{eq5}
\end{equation}
for which the simplest linear solution is the Fourier's law (\ref{eq2}). Although many other solutions exist in the nonlinear regime (i.e., nonlinear functions respect to $\nabla (1/T)$), but we keep our focus on the linear ones. Such linear solution of (\ref{eq4}) still allows to have a state variable-dependent transport coefficient in the constitutive equation such as $\lambda=\lambda(T)$.

In the non-Fourier case, either the set of state variables can be extended, e.g., with the heat flux $\mathbf q$, or the entropy flux $\mathbf J_s$ can be more general \cite{Nyiri91, SzucsEtal20}. Among the various approaches, the non-equilibrium thermodynamics with internal variables (NET-IV) is applied here due to its useful generality \cite{FulEta14m1, BerJurMau11, VanEtal08, MauMus92P1, MauMus92P2}. In other words, while the kinetic theory-based Rational Extended Thermodynamics (RET) \cite{MulRug98, RugSug15} builds its heat conduction models on phonon interactions, the continuum formulation of NET-IV leaves the particular transport mechanism aside. Consequently, in the RET model, the transport coefficients can be calculated, while in the framework of NET-IV, these coefficients can be determined through a fitting procedure. Therefore, while it is easier to determine the transport coefficients in RET, it is restricted to systems with high Knudsen number (Kn $\geq 0.1$), characteristic for nano or low-temperature systems. On the contrary, the NET-IV model has more freedom by having tunable parameters, and that generality allows us to apply the generalized equation for room temperature problems. Particularly for the Guyer-Krumhansl equation, in the NET-IV approach we have $s=s(e, \mathbf q)$ and $\mathbf J_s = \mathbf B \mathbf q$, in which $\mathbf B$ is called current or Nyíri multiplier \cite{Nyiri91}, helping to introduce the so-called nonlocal terms (e.g., the $\partial_{xx} q$). For the detailed derivation, explanation, and background, we refer to \cite{Van01ch}.
The GK equation for a one-dimensional situation in Cartesian coordinates is
\begin{equation}
\tau_q \partial_{t} q + q = - \lambda \partial_x T + \kappa^2 \partial_{xx} q,
\label{eq6}
\end{equation}
where $\kappa^2$ is a sort of `dissipation parameter', usually connected to the mean free path in the kinetic theory. Eq.~(\ref{eq6}) has two special cases. The first one is the Cattaneo-Vernotte (CV) equation (with $\kappa^2=0$), the other is a less-known one, called Nyíri equation (with $\tau_q=0$):
\begin{equation}
q = - \lambda \partial_x T + \kappa^2 \partial_{xx} q,
\label{eq7}
\end{equation}
which is analogous with the Brinkman equation (the nonlocal generalization of Darcy's law). Although the CV equation is well-applicable for dissipative wave propagation (`second sound'), it is not observed for macro-scale heterogeneous materials on room temperature.
At that point, we note that many other procedures are possible, depending on the purpose and on the thermodynamic framework in which the problem is formulated \cite{Cimm09diff, CarloEtal16, Grmela2018b, Grmetal11, JouEtal99, SauEtal20}.

\subsection{Why the Guyer-Krumhansl model?}

\begin{figure}[H]
\centering
\includegraphics[width=8.3cm,height=5.5cm]{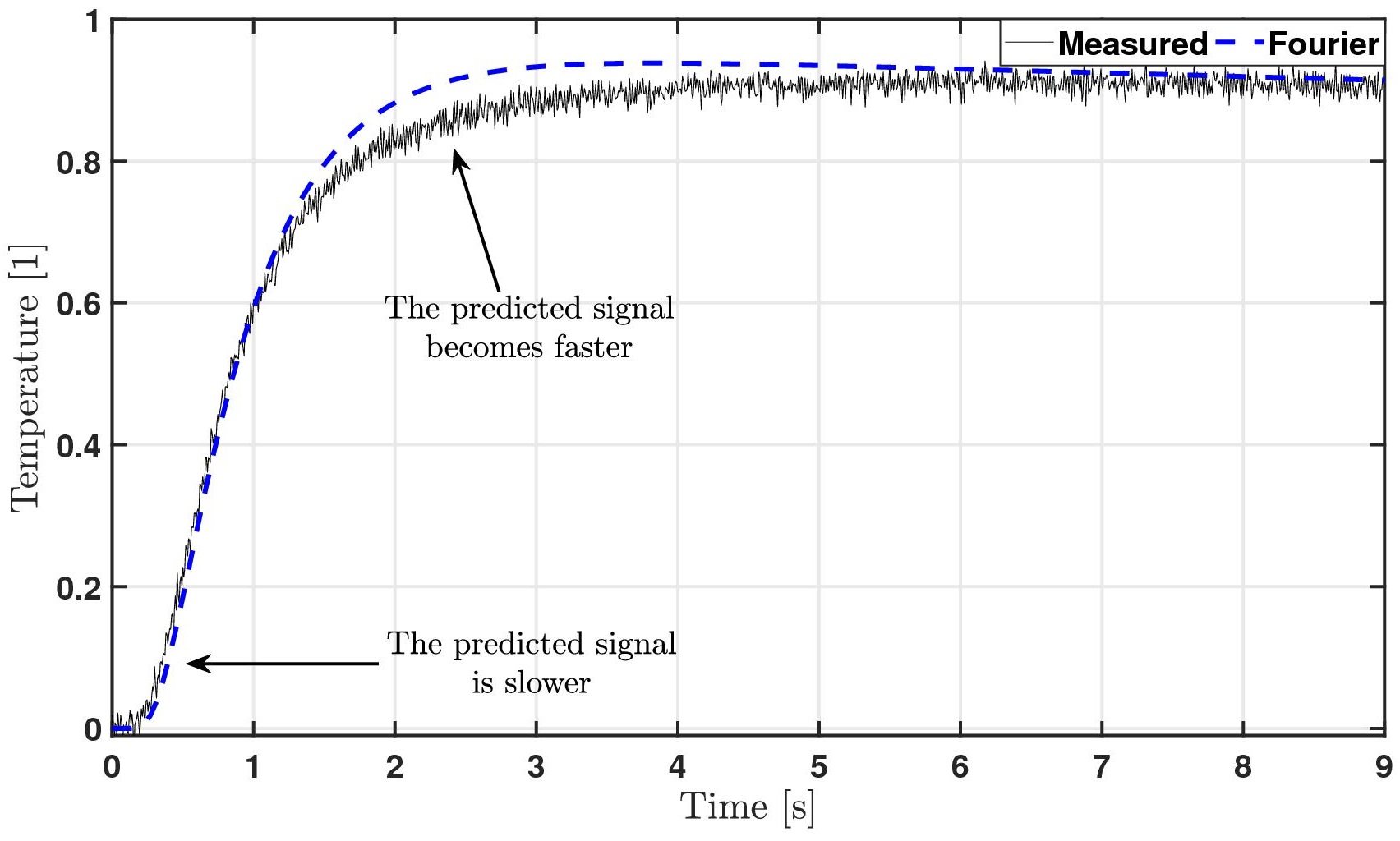}
\caption{The signs of non-Fourier heat conduction for heterogeneous materials at room temperature.}
 \label{fig1}
\end{figure}

In our recent heat pulse experiments, we investigated the transient thermal behavior of several heterogeneous materials such as metal foams, rocks, and capacitor specimen \cite{Botetal16, Vanetal17}. Among them, the capacitor sample has the most regular structure in which good conductor and insulator layers are arranged periodically, two materials with different characteristic thermal time scales. The presence of multiple time scales attributes the heterogeneous materials, in general. In the $T$-representation of the GK model,
\begin{equation}
\tau_q \partial_{tt} T + \partial_t T = \alpha \partial_{xx} T + \kappa^2 \partial_{xxt} T,
\label{eq8}
\end{equation}
the thermal diffusivity $\alpha=\lambda/(\rho c)$ is comparable with the ratio of $\kappa^2/\tau_q$. When $\kappa^2/\tau_q=\alpha$, it is called Fourier resonance, and it recovers the Fourier behavior exactly. Otherwise, non-Fourier solutions appear, and these GK parameters represent the existence of parallel heat conduction time scales. On the contrary to the usual belief that merely wave propagation is possible when non-Fourier effects come into the picture, we found the so-called over-diffusion for which $\kappa^2/\tau_q > \alpha$, and Figure \ref{fig1} shows its characteristics. Remarkably, it seems similar to the classical case, but the Fourier equation cannot describe the data with acceptable accuracy. Moreover, both the thermal diffusivity and the non-Fourier effects can be size-dependent \cite{FulEtal18e}. Overall, based on our experimental findings, the GK equation stands as the next candidate beyond the Fourier equation. Neither of its special cases provided acceptable predictions.

\section{Bioheat models}
Roughly speaking, the related heat conduction models belong to two groups, based on the considered constitutive equation, but all aim to predict the tissue temperature. In the first group, Fourier's law is utilized with various biologically motivated extensions, mostly being source terms in the balance of internal energy. Particular examples are the internal heat transfer by blood perfusion, which appeared first in the Pennes' model \cite{Penn48}. Later, more and more physiological details are tried to be implemented. A few well-known examples are related to the coupled two-temperature Chen-Holmes \cite{ChenHolm80} and the Weinbaum-Jiji-Lemons models \cite{WJL84}. However, the methodology of implementing the detailed structure (e.g., the artery-vein pairs, various tissue structures) does not seem to be advantageous due to the more complicated equations, the unknown variables, and limited validity region.
Also, in some cases, the extensions are physically contradictory: in Wulff's model \cite{Wulff74}, the mean blood velocity $\mathbf u_b$ explicitly appears:
\begin{equation}
\mathbf q = -\lambda_t \nabla T_t + \rho_b f_b \mathbf u_b
\label{eq9}
\end{equation}
with $f_b$ being the blood enthalpy, and the indicies $t$ and $b$ are referring to the tissue and blood properties. Such coupling between $\mathbf q$ and $\mathbf u$ is not possible, unfortunately, due to objectivity reasons \cite{Musch98, Fulop15, MatVan06}. Similar shortcomings appear in Klinger's approach \cite{Kling74, Kling78}.

The second group of biologically motivated heat conduction models consists of the DPL and the CV equations in the vast majority of problems, with keeping specific source terms. Sadly, the DPL model suffers from physical and mathematical shortcomings, and the CV equation predicts waves, which are not observed so far.

Beyond these models, there are more specific ones in the literature. For instance, finding the thermal response of the cornea is a long-lasting question. In this respect, we mention the work of Taflove and Brodwin \cite{TafBro75} due to their remarkable idea: it is more advantageous to use `effective' models. Effective in the sense that the complex heterogeneous structure is substituted with a homogeneous one; consequently, the model parameters are also effective, i.e., `averaging' the complete thermal behavior.
This methodology agrees with the GK model's perspective: it introduces two new, physically strongly motivated parameters, which characterize the entire medium, despite its particular heterogeneities. From a purely thermodynamic point of view, rocks do not differ significantly from biological materials: they both porous, having various heterogeneities and irregularities, thus parallel time scales are present. If such material shows non-Fourier thermal behavior, then it should be observable when all biological specific effects are excluded, and no heat source is present. In other words, the difficulties with complicated heat sources and boundary conditions must be separated from the constitutive model. It stands as the motivation to revisit two preceding experiments in which these aspects appear and could lead to the misinterpretation of the observed data.

\section{Experiment I.: processed meat samples}
\begin{figure}[H]
\centering
\includegraphics[width=6.3cm,height=11.5cm]{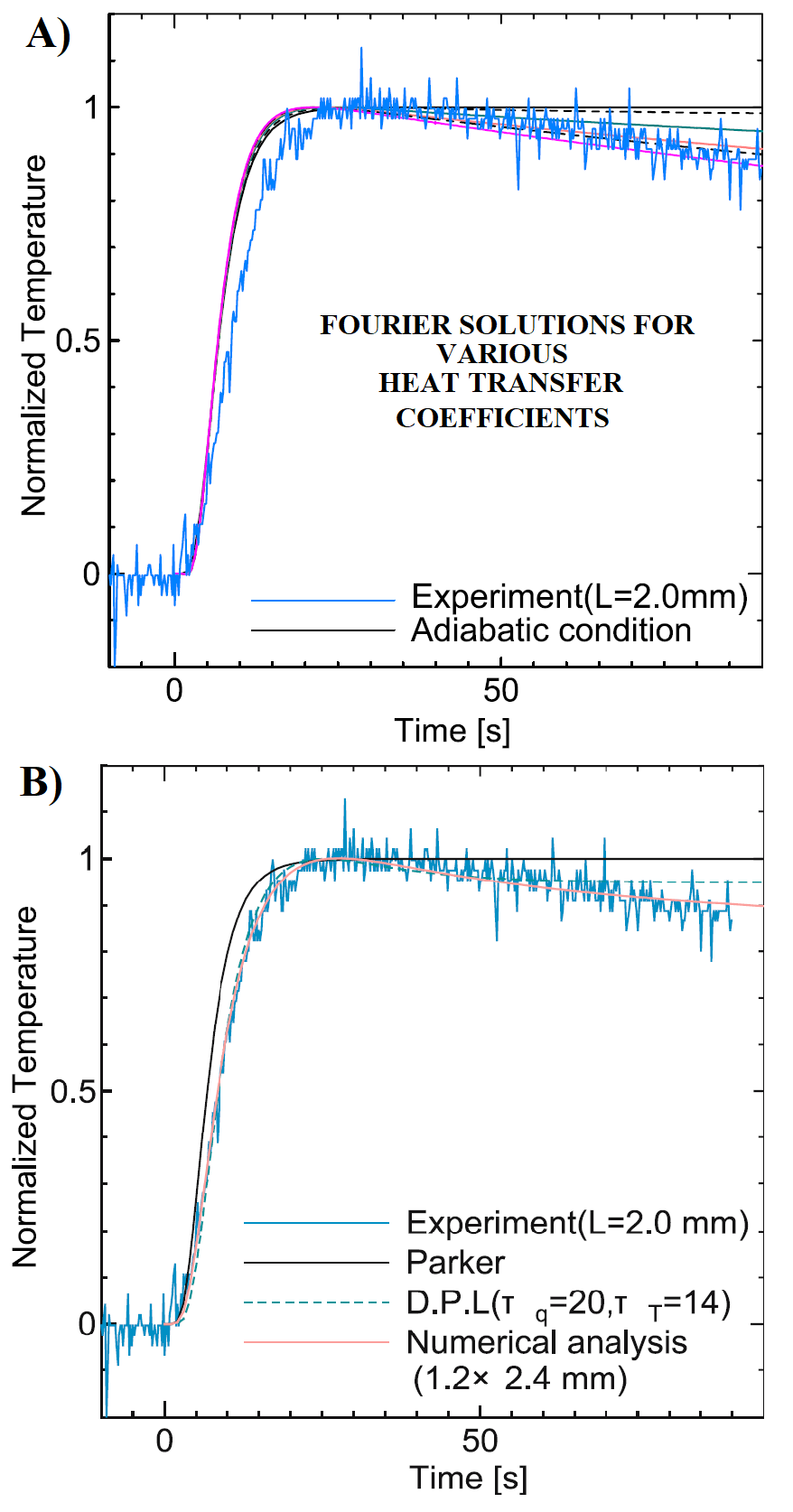}
\caption{The observed temperature history for the thinnest sample ($2$ mm). A) presents the solutions for Fourier equation with various heat transfer coefficients. B) shows the outcomes of other approaches. The complete figures and for further information, we refer to \cite{TanEtal07}.}
\label{fig2}
\end{figure}

The first experiment we re-evaluate is performed by Tang et al.~\cite{TanEtal07} on processed meat samples, including fat as inclusions. They prepared three samples with $2$, $3$ and $4$ mm thickness, and all are having the same $10$ mm diameter. They applied a heat pulse on the front side uniformly, and it lasts $1$ s while they registered the rear side temperature history for $90$ s. Figure \ref{fig2} shows their results partially; for their complete investigation, we refer to \cite{TanEtal07}. According to their interpretation, convection cannot be responsible for the measured deviation from Fourier's law. They investigated two approaches to explain this phenomenon. On one hand, they proposed to use the DPL model with adiabatic (zero flux) boundaries, hence neglecting the possible heat transfer to the environment. On the other hand, they introduced the size of the fat inclusion as a variable parameter and performed a detailed numerical analysis using the Fourier equation.

In the end, they concluded that the Fourier equation seems to be a better choice together with these new parameters than the DPL model. These models differ mostly at the end of the temperature history. Despite the valuable insight they provided by these experiments, there are a few misinterpretations, which we found to be important presenting here.

We re-evaluate that data using a novel algorithm developed explicitly for the GK equation using its analytical solution \cite{FehKov21}. Shortly, this evaluation procedure first finds the heat transfer coefficient and the thermal diffusivity. If needed, it continues to find the GK parameters ($\tau_q$ and $\kappa^2$). We found that omitting the heat transfer to the environment yields significant errors in the modeling, i.e., Fourier's law and convection boundary condition provide a suitable model to adequately describe the observed data. In other words, no generalized heat conduction equation is needed, merely a proper optimization on the effective thermal diffusivity. Consequently, despite that they found the relaxation times in the mathematically valid domain ($\tau_q>\tau_T$), it remains misleading to interpret the temperature history as proof for the wave nature of heat conduction under such conditions.

\begin{figure}[H]
\centering
\includegraphics[width=8.3cm,height=5.5cm]{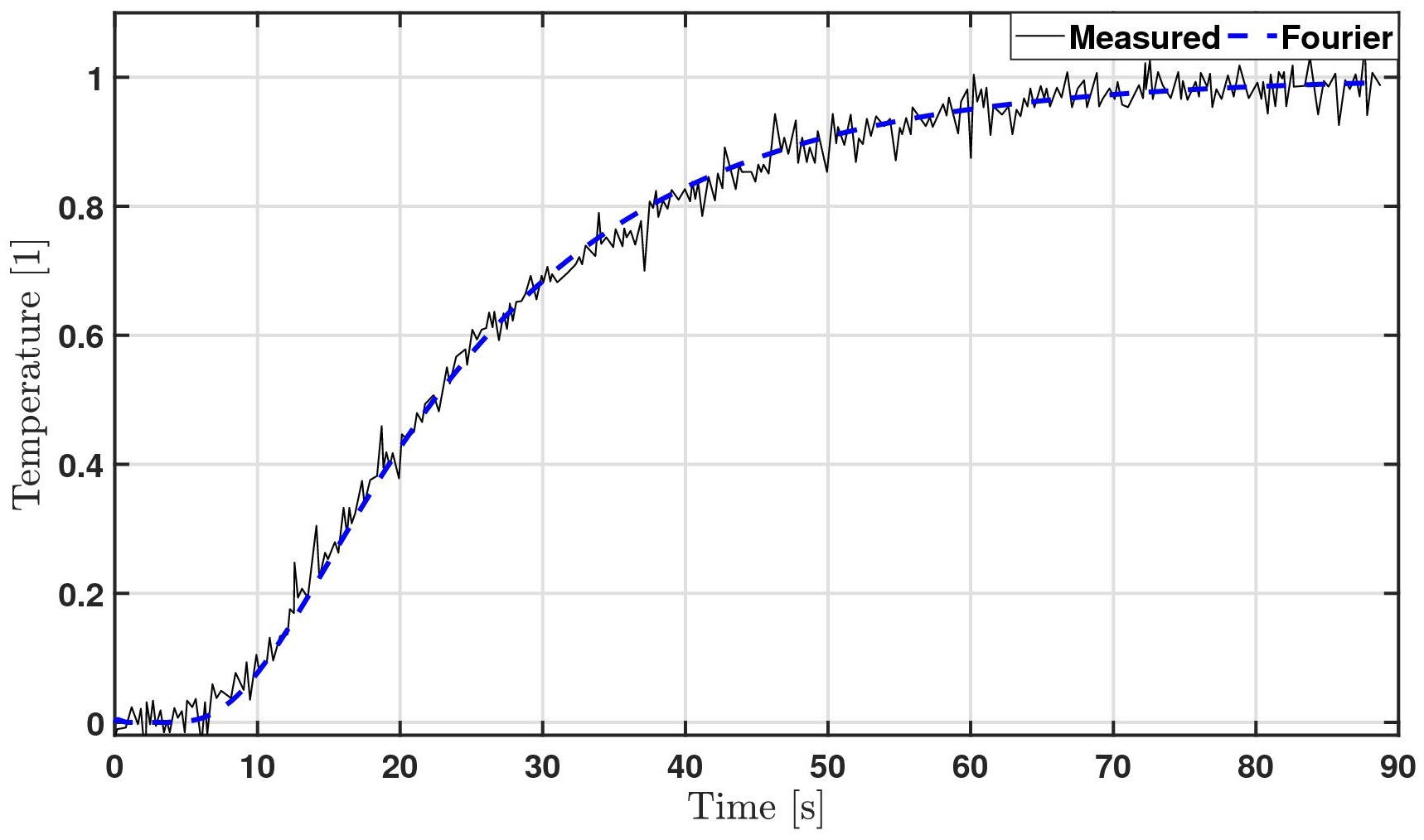}
\caption{Comparing the Fourier model to the sample with $4$ mm thickness. The measured data is recovered using a plot digitizer.}
 \label{fig3}
\end{figure}

\begin{figure}[H]
\centering
\includegraphics[width=8.3cm,height=5.5cm]{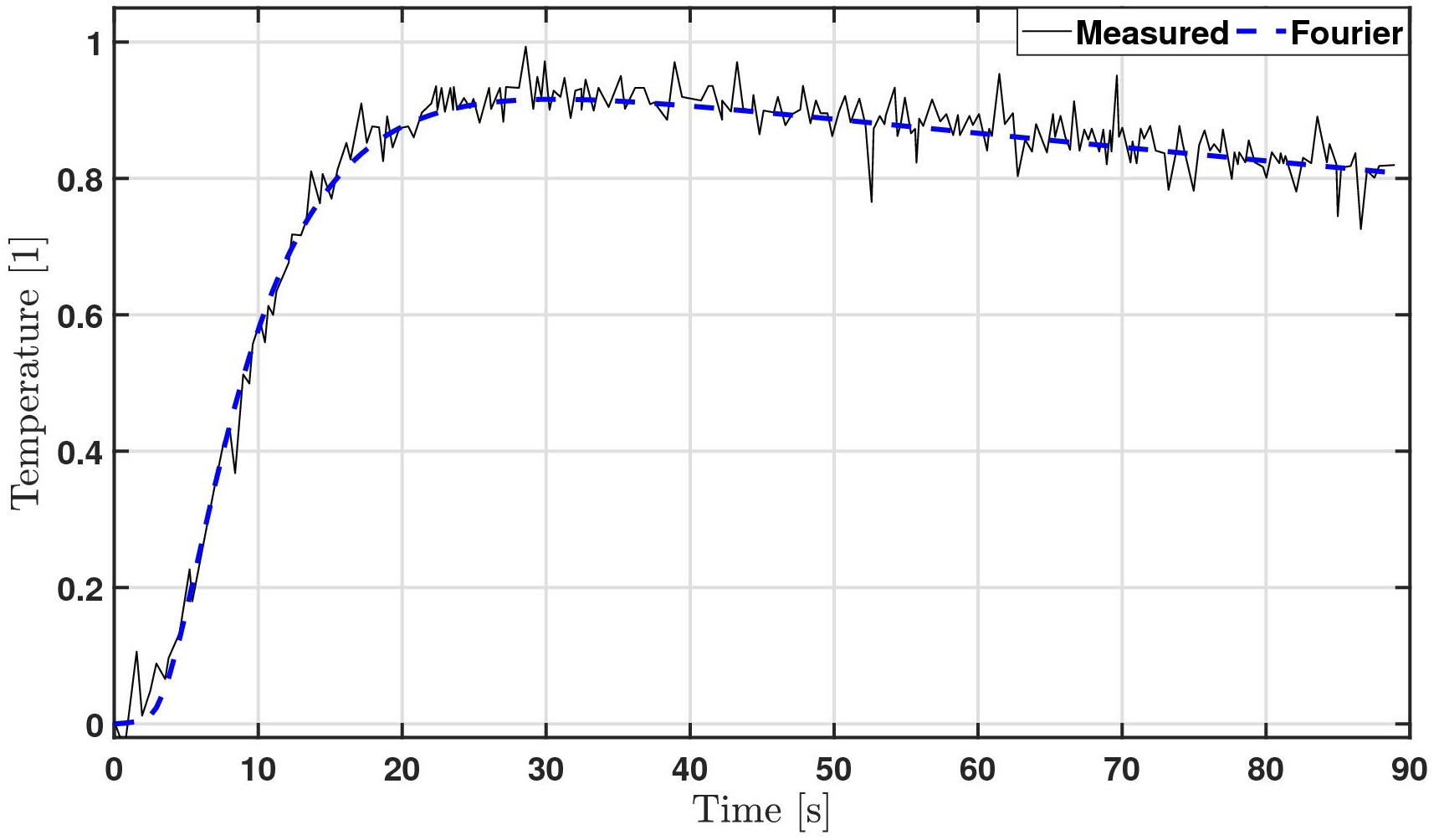}
\caption{Comparing the Fourier model to the sample with $2$ mm thickness. The measured data is recovered using a plot digitizer.}
 \label{fig4}
\end{figure}

Nevertheless, the size-dependent behavior is indeed present within this spatial scale ($2-4$ mm) and requires further investigation, and probably new experiments would be helpful for the literature. Table \ref{T1} summarizes the parameters, and Figures \ref{fig3}-\ref{fig4} show the fitting of the Fourier equation.

\begin{table}[h]
\centering
\begin{tabular}{cc}
\multicolumn{1}{l}{}                  & \multicolumn{1}{l}{}                                                              \\
\multicolumn{1}{c|}{Sample thickness} & \begin{tabular}[c]{@{}c@{}}Thermal diffusivity\\ $10^{-7}$ [m$^2$/s]\end{tabular} \\ \hline
\multicolumn{1}{c|}{$2$ mm}           & $0.66$                                                                            \\ \hline
\multicolumn{1}{c|}{$3$ mm}           & $0.73$                                                                            \\ \hline
\multicolumn{1}{c|}{$4$ mm}           & $1.04$
\end{tabular}
\caption{The found thermal diffusivities with the Fourier equation.}
\label{T1}
\end{table}

\section{Experiment II.: skin tissue samples}
In this section, we revisit the experiment of Jaunich et al.~\cite{Jaetal08}, in which they prepared various cylindrical samples made from the mixture of araldite, DDSA resin, and DMP-30 epoxies in order to mimic the behavior of real tissue, moreover, they added titanium dioxide to this mixture as a `scatterer' for the laser light. They assumed that this mixture also thermally substitutes the real skin tissue, therefore utilizing the same optical and thermal parameters, detailed in \cite{Jaetal08}. A skin tissue has a promising structure in regard to non-Fourier heat conduction due to the parallel time scales (the presence of various layers and heat transfer mechanisms). Nevertheless, its observation might also require the heat flux to be parallel with these layers \cite{Botetal16}. Contrary to the previous experiment, the thermal excitation now lasts $10$ s instead of $1$ s, and registering the temperature history for $10$ s.

Regarding the aspects of non-Fourier heat conduction, all these are extremely important. The material structure has no sharp interfaces inside due to the relatively homogenous mixture of epoxy components, therefore it is different from real heterogeneous porous materials. Furthermore, the excitation is unusual in the sense that it is not `pulse-like' but much longer, significantly absorbed under the surface, and the temperature history is measured only in the period of laser heating, and there is no available data after. Consequently, the entire experiment is heat source dominated, which hides heat conduction and, eventually, almost any other thermal transport effects.

\begin{figure}[H]
\centering
\includegraphics[width=7.5cm,height=14cm]{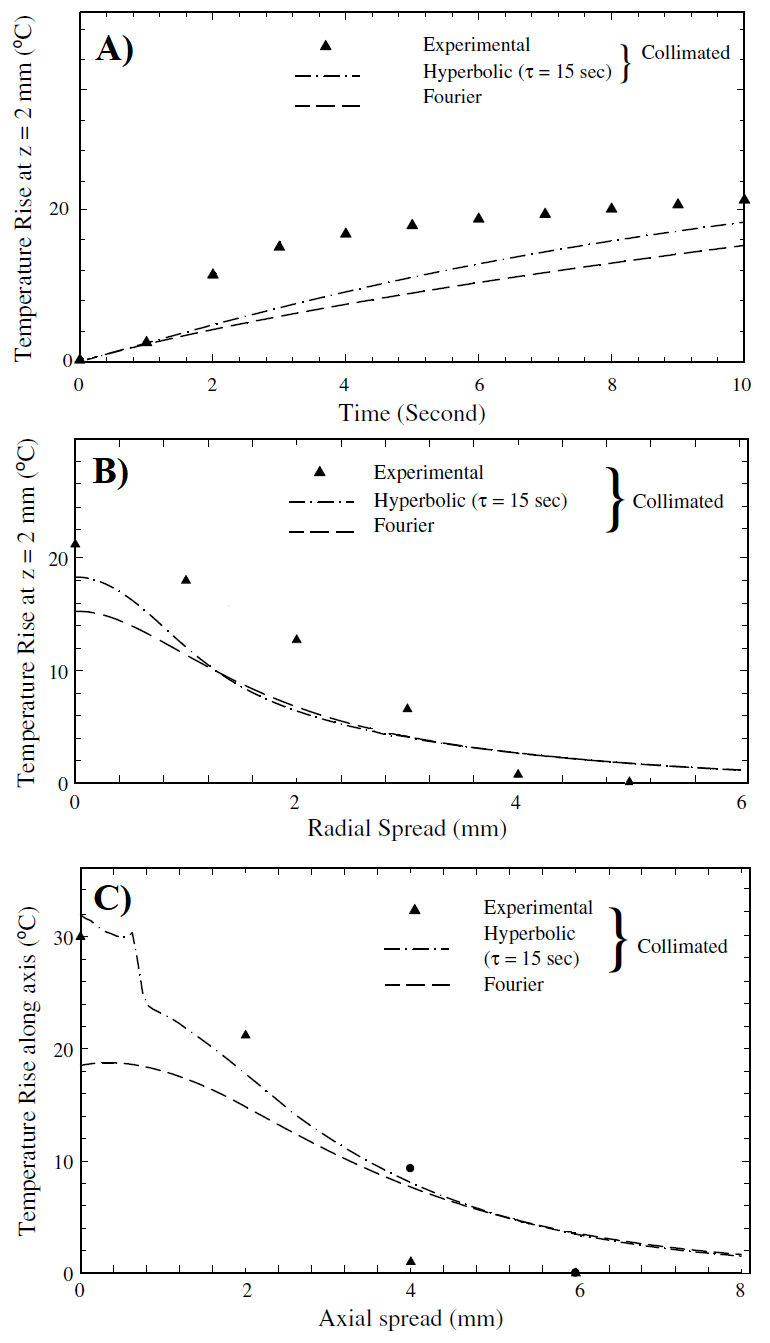}
\caption{The partial results of Jaunich et al., for their detailed description and for the complete figures, we refer to \cite{Jaetal08}. Triangles denote the measured temperature increase. A) presents the temperature history at $2$ mm depth on the axis of the cylinder. B) and C) show the radial and axial temperature distribution at specific points.}
 \label{fig5}
\end{figure}

According to their findings, Fourier's law is seemingly inadequate for this problem since it predicts the temperature field quite far from the measured one. Figure \ref{fig5} presents the measured and fitted results partially, only the data, which we also tried to reproduce. The assumption that other constitutive equations should be used instead of Fourier's law requires accurate knowledge of any other influencing property, most importantly, on the internal heat generation caused by the laser light. In their study, the CV (hyperbolic) equation is tested and found to be better than the Fourier equation, providing more accurate predictions. Unfortunately, even the CV equation fails to offer the proper values, and at some points, it is still the same as the Fourier equation.

\begin{figure}[H]
\centering
\includegraphics[width=8.5cm,height=5.5cm]{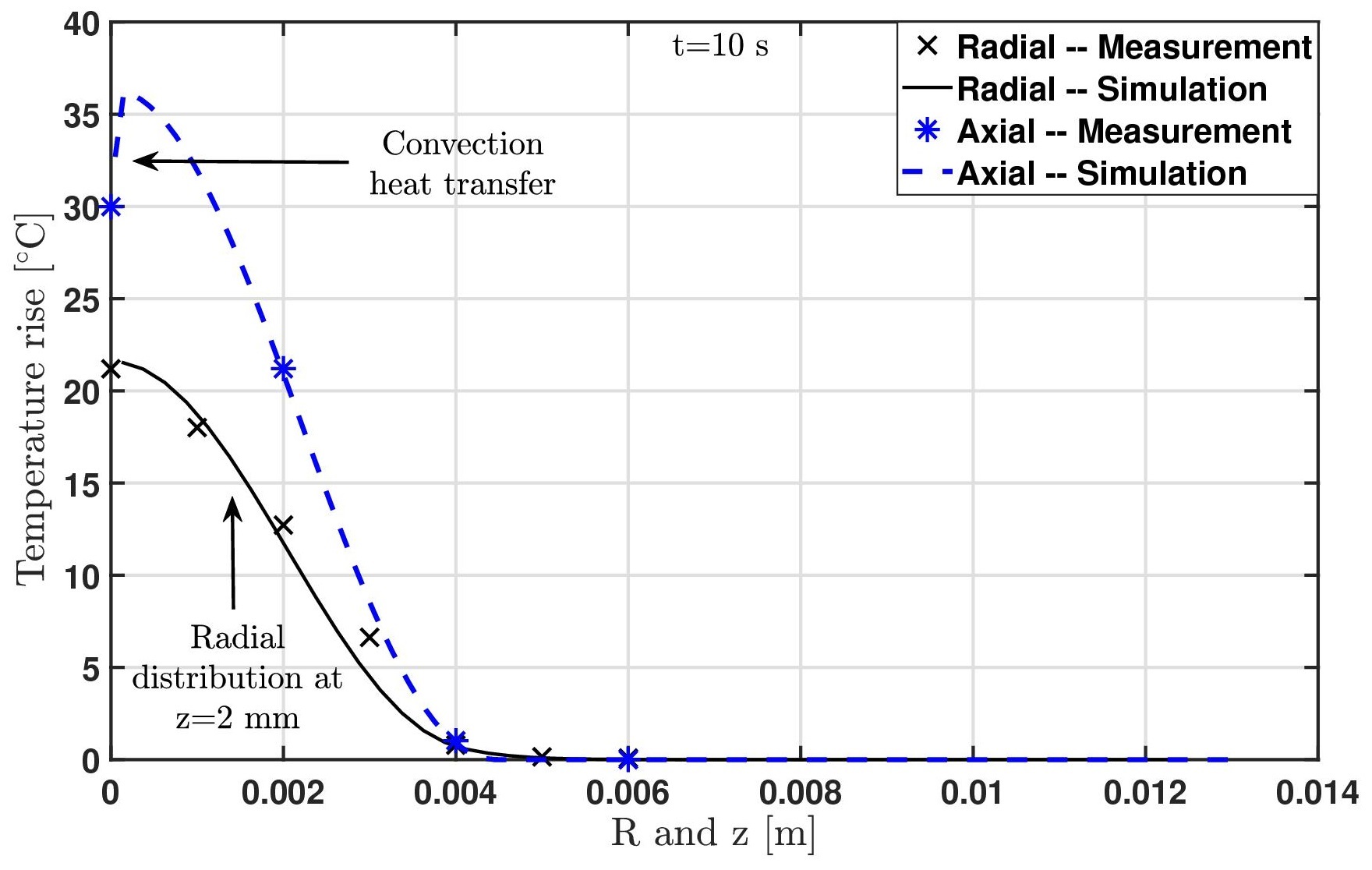}
\caption{Comparing the Fourier model to the measured values in both axial and radial directions. The measured data is recovered using a plot digitizer.}
 \label{fig6}
\end{figure}

\begin{figure}[H]
\centering
\includegraphics[width=8.5cm,height=5.5cm]{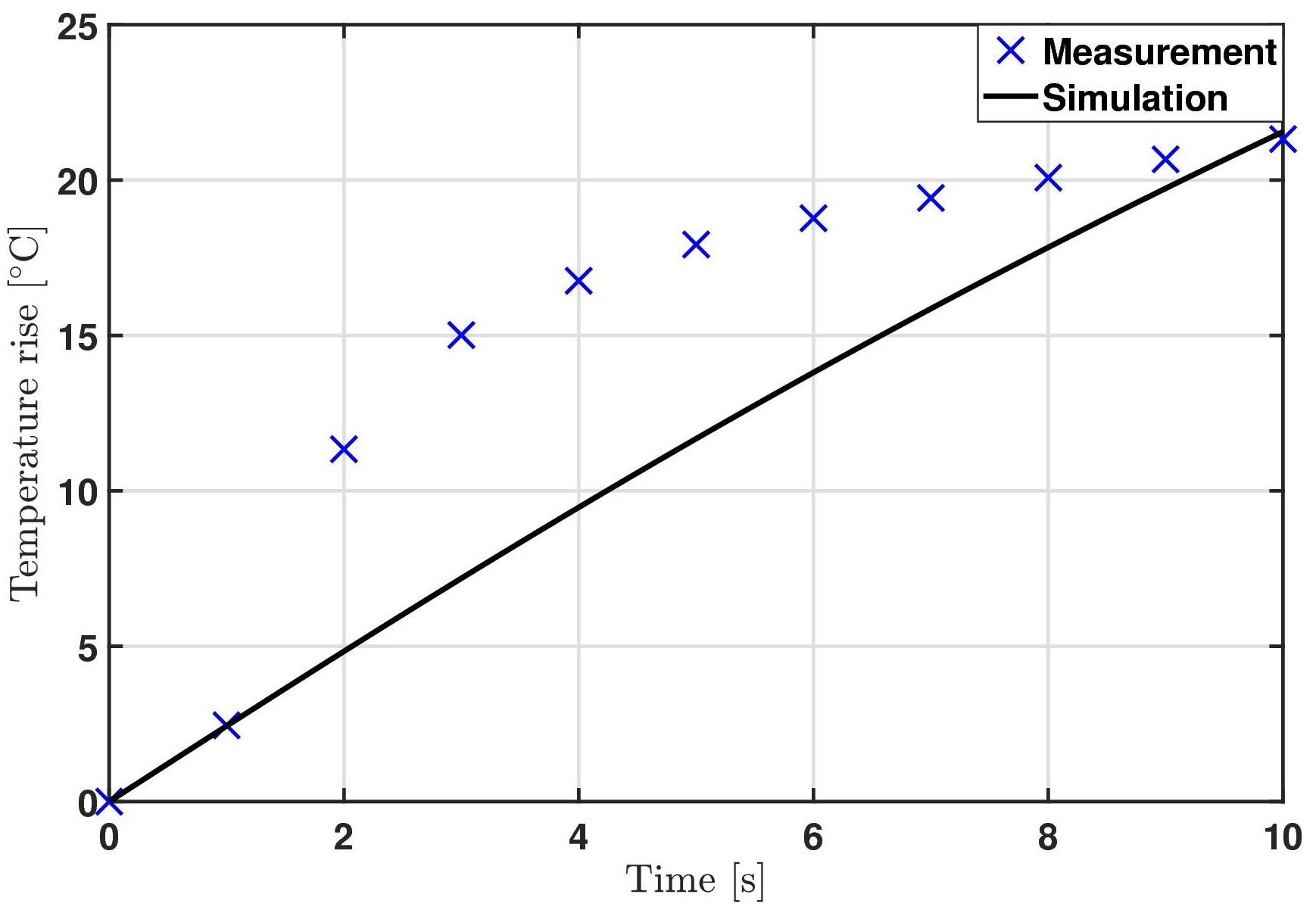}
\caption{Comparing the Fourier model to the temperature history recorded at $z=2$ mm. The measured data is recovered using a plot digitizer.}
 \label{fig7}
\end{figure}

We also tried to model this experiment by solving the Fourier, CV, and GK equations in cylindrical coordinates based on the scheme published in \cite{JozsKov20b}. We found that the problem indeed originates in the heat source since neither the CV nor the GK model could improve the fitting, as it was supposed from the beginning. Therefore we tried various volumetric heat generation models from which the following one provides the closest values with the Fourier equation:
\begin{equation}
Q_v=\left \{ \begin{array}{ll}
\displaystyle
\frac{q_m}{2} \left(1+\cos\big(\pi r_e /r_p \big) \right) \ & \ r_{{e}}<r_{{p}} \\
0 \ & \ \textrm{else}
\end{array} \right.,
\end{equation}
in which
\begin{equation}
r_e=\sqrt{r^2 + (z-z_c)^2}
\end{equation}
is the effective radius where heat generation $Q_v$ occurs, $r_p$ stands for the pulse radius; $z_c$ appears in the case when the center of heat generation is under the surface, in our case, it is found to be $z_c=0$; $q_m$ is the maximum heat generation. The value of $q_m$ is assumed to be the only adjustable parameter due to the unknown amount of absorbed heat, and it is found to be $q_m=1.7\cdot 10^{7}$ W/m$^3$, which means that around the half of the irradiated energy is turned into heat generation. Furthermore, $z$ and $r$ are the axial and radial coordinates, respectively. Although this is an empirical model, it is motivated by statistical physics: photons propagate with equal probability in every direction after enough number of scattering, resulting in a spherically symmetric absorption. Figures \ref{fig6} and \ref{fig7} present the best outcome we managed to achieve.

As it is apparent from Fig.~\ref{fig6}, we also applied a convection boundary on the surface, which was irradiated by the laser pulse, the heat transfer coefficient is assumed to be $10$ W/(m$^2$K). Unfortunately, the measured transient characteristic at $z=2$ mm is still uncovered, and non-Fourier models cannot offer a better description, in agreement with \cite{Jaetal08}.

\section{Discussion}
The thermal modeling of biological materials is a challenging task and has increasing attention due to its numerous applications. In the present paper, we aimed to offer a brief overview of the essential aspects of non-Fourier equations in continuum thermodynamics, focusing on their observability in room temperature experiments.

First, we want to emphasize that one must pursue to separate the heat sources from the heat conduction effects as much as possible. Although volumetric heat generation is inevitable in many practical situations and also has a significant influence in medical diagnostics \cite{SudarEtal20a, SudarEtal20b, AndrEtal19}, it could dominate the time evolution of the temperature field. Therefore, the experiments, which aim to measure biological material's thermal properties, should be designed to let the heat conduction be the dominant heat transport mechanism instead of including complex heat source and heat generation mechanisms. In such a preferable situation, it is easier to decide whether we observed a non-Fourier phenomenon or not; consequently, the role of a constitutive equation becomes apparent in the thermal model. Moreover, the non-Fourier thermal parameters are more comfortable to determine. Otherwise, one can easily misinterpret the measurements and concluding in a different outcome, and it is recommended to always test all the possibilities in a model before implementing a generalized constitutive equation.

One must be careful since the observation of non-Fourier propagation depends both on the boundary conditions and material properties. In other words, it requires the coexistence of suitable spatial and time scales, probably having a size-dependent behavior as well. This field is still under development, and there are countless questions, which wait to be solved in the future. Also, as our re-evaluation attempts are standing here, the Fourier equation can explain most of the observations, and most of the seeming deviation disappeared.

Overall, we aim to collect and call attention to these modeling aspects. The research field of non-Fourier heat conduction is vast, continuously growing, and increasingly challenging to keep up with the up-to-date results, mainly because these constitutive equations require a deeper understanding of boundary conditions as well. The classical definitions established for the Fourier equation do not work, and even for a thermodynamically strongly motivated model (such as the GK equation) can lead to unphysical, false solutions with inadequate boundary conditions. Therefore, despite the attractive results of non-equilibrium thermodynamics, it is suggested to be careful with generalized models.

\section*{Acknowledgement}
We dedicate our paper to the memory of our respected colleague, Prof. José Casas-Vázquez, who passed away recently. \\ The research reported in this paper and carried out at BME has been supported by the grants National
Research, Development and Innovation Office-NKFIH FK 134277, and by the NRDI Fund (TKP2020 NC, Grant No. BME-NC) based on the charter of bolster issued by the NRDI Office under the auspices of the Ministry for Innovation and Technology.
This paper was supported by the János Bolyai Research Scholarship of the Hungarian Academy of Sciences.



\end{document}